\documentclass{ejaps_arXiv}
\usepackage{graphicx}
\usepackage{amsmath}
\usepackage{xcolor} 
\usepackage{CJKutf8}

\usepackage[whole]{bxcjkjatype}

\def\changed#1{{#1}}

\begin{document}
\Year{2025}
\Vol{xx}
\No{x}
\title{Prigogine's temporalization of physics: two agnostic attitudes of physicists}
%\title[title for header]%
%      {title}
%\subtitle{}
\authorlist{%
 \authorentry{Hirokazu Maruoka}{a}
}
%\breakauthorline{3}
%\headauthorlist{}
%\affiliation{}
\affiliate[a]{Nonlinear and Non-equilibrium Physics Unit, Okinawa Institute of Science and Technology (OIST), Tancha, Onna-son, Kunigami-gun Okinawa 904-0495,
Japan}{hirokazu.maruoka@oist.jp, hmaruoka1987@gmail.com}
%\ack{This work has been supported ... etc.}

\begin{abstract}
In this paper, I aim to clarify the unconscious ideologies and attitudes held by physicists through Prigogine's work. Prigogine was an outstanding chemist and physicist who made significant contributions to the development of non-equilibrium thermodynamics. At the same time, he extended his ideas beyond physics into the humanities, engaging in an interdisciplinary exploration of scientific and philosophical thought. Due to his unique career, Prigogine's reception has been deeply divided.  This study highlights his intellectual endeavors to formulate two distinct agnostic attitudes---one held by dynamicists and the other by Prigogine himself. Building on this formulation, I examine how physics has been spatialized, drawing on the philosophy of Bergson. Finally, I explore an alternative path that Prigogine might have envisioned---the temporalization of physics---in a broader context to extend and revitalize his philosophy.
\end{abstract}
\begin{keywords}
Prigogine, Physics, agnosticism, spatialization, temporalization
\end{keywords}
%\received{}{}{}
%\revised{}{}{}
%\accepted{}{}{}
\maketitle

\section{Introduction}%

Prigogine was a pioneering figure in the development of non-equilibrium thermodynamics in the postwar period, and his contributions earned him the 1977 Nobel Prize in Chemistry \cite{Nobel,Maruani2004}. As a scientist, he played a central role in the foundation and development of non-equilibrium theory. However, he was also known for his intellectual engagement with the humanities, maintaining dialogues with scholars outside of physics and publishing works that bridged natural science and the humanities.

In a study by Kondepudi et al. \cite{Kondepudi2017}, Prigogine's research is categorized into three distinct phases: first phase –  "the early period in which he reformulated thermodynamics as a science of irreversible processes, thus changing it from a theory of states as it was formulated in the 19th century". Influenced by Onsager's reciprocal relations \cite{Onsager1931}, Prigogine sought to generalize classical thermodynamics to include non-equilibrium phenomena. This effort culminated in his 1947 doctoral dissertation \cite{Prigogine1947}. Second phase – "the period in which his Brussels-Austin group formulated the theory of dissipative structures and conducted extensive studies of diverse systems". It is the expansion of non-equilibrium thermodynamics, particularly in linear systems, followed by the exploration of self-organization in nonlinear systems. The Brussels School, led by Prigogine, played a major role in this development. After Prigogine was appointed as a professor at the University of Texas at Austin, the Brussels-Austin group introduced the concept of {\it dissipative structures} \cite{Prigogine1971,Prigogine1977}, which became a key framework for understanding self-organization far from equilibrium. His contributions to non-equilibrium thermodynamics and dissipative structures earned him the Nobel Prize in Chemistry in 1977.

For most physicists, Prigogine's achievements in the first and second phases are widely accepted, with little controversy regarding their significance \footnote{Some litterateurs are critical to the second phase, such as Ref. \cite{Oono2013}.}. Indeed, his Nobel Prize was awarded specifically for his contributions during these two periods. However, his third phase is particularly notable. Third phase – "the later period in which his group focused on making irreversibility a fundamental part of physics by extending the formulation of mechanics", Prigogine published broadly interdisciplinary and popular works, such as From Being to Becoming \cite{Prigogine1980}, Order Out of Chaos \cite{Prigogine1984}, and the End of Certainty \cite{Prigogine1997} engaging not only with physics but also with humanities, sociology, and philosophy in collaboration with Stengers \cite{Stengers,Belyi}. Prigogine's works from this phase received a mixed reception. While philosophers and humanities scholars often viewed his work positively \cite{Gunter}, as it aligned with contemporary intellectual movements, physicists were far more skeptical \cite{Bricmont1995,Tasaki1997}. His broad claims extending into philosophy were particularly criticized, and his mathematical transformation theory, which sought to integrate irreversibility into mechanics, failed to gain widespread acceptance in contemporary physics. Thus, Prigogine's legacy is deeply divided, with his contributions in non-equilibrium thermodynamics widely recognized, while his later philosophical and theoretical efforts remain contentious.

When assessing Prigogine's impact, it is crucial to distinguish between his scientific contributions and his underlying intentions. From a technical point of view, his third-phase research was not widely embraced by the physics community, and some physicists, such as Bricmont \cite{Bricmont1995}, have provided detailed critiques of his work. However, these scientific criticisms often fail to fully engage with the philosophical ideas that Prigogine sought to develop during this period. \changed{They tend to emphasize the technical aspects of the theory while overlooking its underlying philosophical motivation.} Prigogine's third-phase philosophy was deeply intertwined with the results of his first and second phases, yet its roots can be traced back to his intellectual concerns even before he formally entered physics. His philosophy extends beyond physics, reaching into the unconscious worldviews that physicists themselves hold about nature. By reconsidering Prigogine's philosophy, we can illuminate the implicit assumptions and worldviews that shape modern physics. Nonetheless, the philosophy implicitly suggested therein appears insufficiently developed by Prigogine himself and has generally been regarded as secondary to his primary scientific achievements.

This study aims to clarify the strategic and conceptual foundations of time in physics through the lens of the philosophy of Prigogine. In his philosophy, one can discern a distinctive stance towards physics and nature, which I attempt to formulate as a kind of {\it agnosticism}. Building on this formulation, I aim to extend and reinterpret Prigogine's notion of the {\it temporalization of physics}, which is a fundamentally different way of physics.

The structure of this manuscript will be as follows. First, by building on Prigogine's theoretical framework, we demonstrate that physics reveals four distinct concepts of time, each of which is classified by certain attributes (Sections 2 and 3). This reveals the underlying tendencies and strategies of physics regarding time. Second, we show that these concepts of time are inseparably linked to the worldview of physicists, particularly to the dynamical perspective of mechanics (Section 4). In particular, we highlight how dynamics and thermodynamics differ fundamentally in their treatment of dissipation. Third, we analyze the implicit assumptions held by dynamicists and contrast them with Prigogine's critique, ultimately identifying a fundamental epistemological attitude (Section 5). This contrast reveals two fundamentally different agnostic attitudes toward knowledge. Finally, we find that these two attitudes give rise to two different ways of physics, the one is the spatialization of physics (section 6) and temporalization of physics (section 7). Through a path of the philosophy of Prigogine, we would like to discuss the time of nature in physics now and in the future.

\section{Concept of time in Physics I: the physics of being and the physics of becoming} %

%A retrospective examination of Prigogine's research reveals that he fundamentally approached natural science from a humanities-oriented perspective. This marks a significant contrast to scientists who, having been deeply immersed in natural science from the beginning, later extend their perspective toward philosophical considerations from within the scientific domain itself. Prigogine, however, was able to observe natural science from an external vantage point, and this perspective is deeply reflected in his critique of physics.

In this section, we attempt to clarify the concept of time in physics by first examining Prigogine's classification of time concepts. Prigogine proposed a fundamental bifurcation of physical theories based on whether their concept of time is reversible or irreversible. This idea was later expanded in his work From Being to Becoming \cite{Prigogine1980}, where he classified physical theories into two major frameworks: the {\it physics of being}, which is founded on a reversible concept of time, the {\it physics of becoming}, which is based on an irreversible concept of time.

The physics of being is defined by time-reversal symmetry, meaning that its fundamental equations remain unchanged under the transformation $t \rightarrow -t$ where \changed{$t$ is time}. This category includes dynamical systems, quantum mechanics, and relativity theory. For instance, a fundamental equation in dynamical systems, the Lagrange equation\cite{Landau}, remains invariant under the transformation $t \rightarrow -t$ as follows,
\begin{equation}
\frac{d}{d\left(-t\right)} \frac{\partial L}{\partial \left(-v\right)} - \frac{\partial L}{\partial r} =\frac{d}{dt} \frac{\partial L}{\partial v} - \frac{\partial L}{\partial r} 
 \label{eq:Lagrange}
\end{equation}
\changed{where $L$ is Lagrangian, $r$ is a configuration and $v = \frac{dr}{dt}$ is a velocity.}

In equations of motion, the solution curves are fully determined once an initial condition (and sometimes a boundary condition) is specified. These solution curves define the trajectory of a system---for $t>0$, they describe the future evolution of the system, while for $t<0$, they describe its past evolution. In fundamental equations that exhibit time-reversal symmetry, there is no intrinsic distinction between motion in the forward direction of time and motion in the reverse direction. This means that whether we observe the motion in forward playback or in reverse playback, no physical distinction exists. The direction of motion is not inherent to the motion itself---it does not influence the fundamental nature of the process. Such motion is classified as a reversible process.

In contrast, everyday experience tells us that some processes inherently contain a preferred direction of time. For example, if we consider the initial condition of a drop of ink falling into a liquid surface, its future evolution can only be one where the ink disperses over time. Karl Popper made a similar observation regarding ripples spreading outward when a drop of water falls into a liquid surface \cite{Popper}. In such cases, the direction of motion is embedded within the process itself. These types of motion exhibit a fundamental distinction between forward playback and reverse playback, making the two directions physically and fundamentally different. This class of motion is referred to as an irreversible process, and the domain of physics that studies such motion is what Prigogine termed the physics of becoming.

The physics of becoming, which is fundamentally based on the concept of irreversible time, is primarily governed by thermodynamics. While thermodynamics shares fundamental principles such as dynamics and energy conservation with classical mechanics, it also introduces a unique concept that is absent in dynamical theory: entropy. Entropy plays a crucial role in distinguishing between reversible and irreversible motion. In thermodynamics, which classifies energy into work and heat, entropy characterizes the constraints governing their exchange. Thermodynamics fundamentally examines the interaction between a system and its external environment, focusing on the energy exchange between them. Entropy consists of two distinct components as follows\cite{Kondepudi1998},
\begin{equation}
d S = d_\mathrm{e} S + d_\mathrm{i} S.
\label{eq:entropy}
\end{equation}
where entropy exchange $d_\mathrm{e}S$, which arises from the exchange between the system and its environment, entropy production $d_\mathrm{i}S$, which is generated intrinsically within the system itself. In a heat engine, which extracts heat from a thermal reservoir and converts it into work, the maximum efficiency of this energy conversion is determined by a single-valued function of temperature. Under ideal conditions, the conserved quantity associated with this exchange corresponds to entropy exchange $d_\mathrm{e}S$, and such a process is classified as a reversible process. However, in real-world interactions, achieving this maximum efficiency is often impossible. The discrepancy between ideal efficiency and actual performance introduces an unknown quantity---dissipation, which corresponds to entropy production $d_\mathrm{i}S$. When $d_\mathrm{i}S$ is generated ($d_\mathrm{i}S >0$), the process becomes irreversible. In contrast, $d_\mathrm{i}S=0$ means that the process is reversible.

The second law of thermodynamics states that $d_\mathrm{i}S \geq 0$, which can be regarded as a theorem asserting the existence of irreversible processes \cite{Planck1908}. Thus, the second law fundamentally serves as the principle that distinguishes reversible from irreversible processes. Entropy plays a crucial role in this distinction: it separates reversible and irreversible processes, and because irreversibility implies that a process cannot be reversed, its change is unidirectional. This means that motion and direction are inseparable in an irreversible process. While dynamics indeed describes the past, present, and future, it treats motion independently of its direction, giving it a spatialized character. For this reason, it is referred to as the {\it physics of being}. In contrast, thermodynamics, which inherently integrates irreversibility and the inseparability of motion and direction, is referred to as the {\it physics of becoming}.

\section{Concept of time in Physics II: the time with or without process}

The distinction between the physics of being and the physics of becoming is fundamentally based on the axis of time reversibility versus irreversibility, which ultimately arises from the presence or absence of \changed{entropy production}. \changed{Prigogine has stressed this distinction based on the time-reversibility. However, to overview the theoretical framework of physics, we easily see that we can further refine the concept of time in physics by examining how change is described. Here we would like to introduce another distinction based on the process of change}.

%%%%%%%%%%%%%%%%%%%%%%%%%%%%%%%
\begin{figure}[h]
\includegraphics[width=12cm]{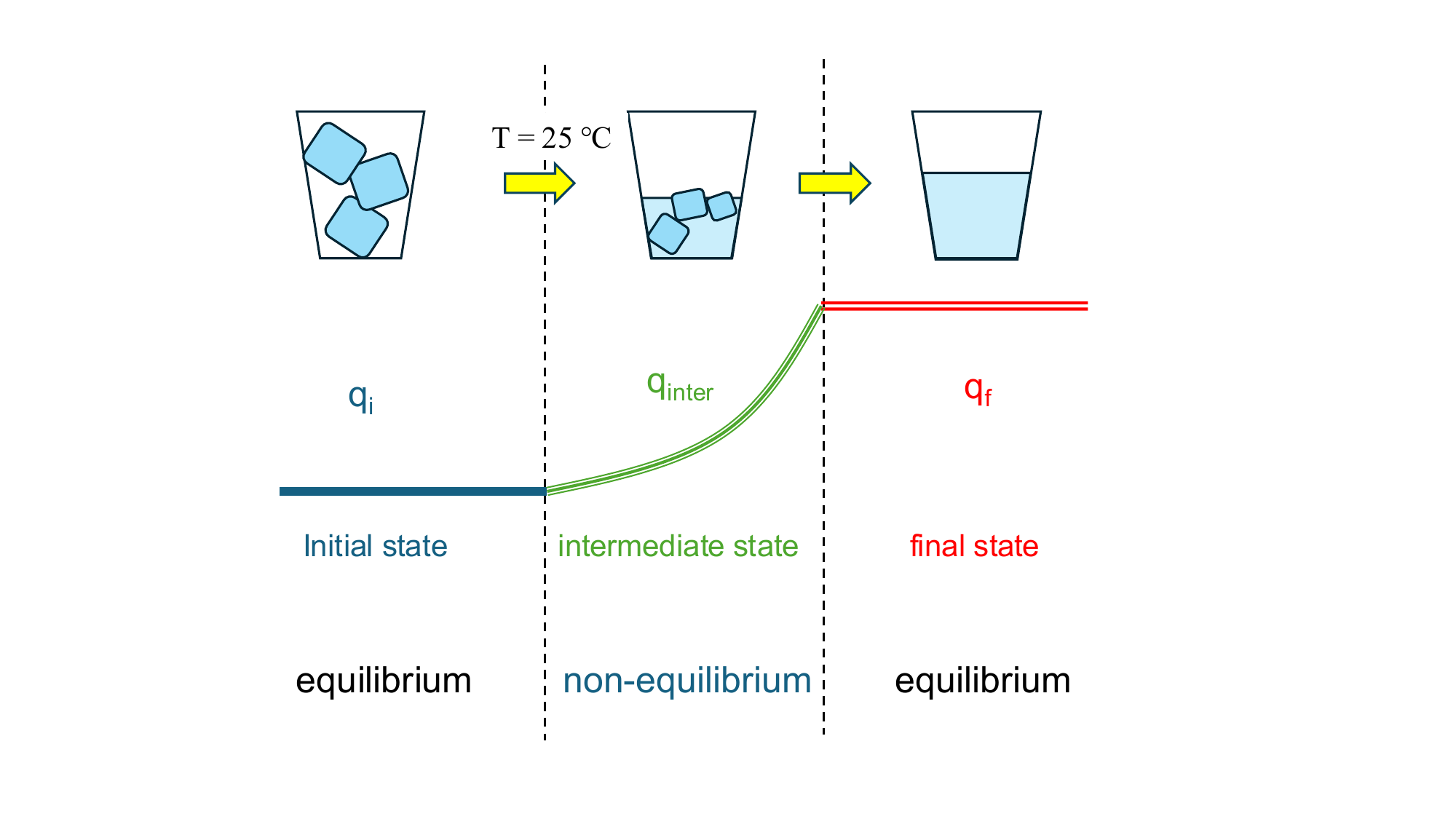}
\caption{The formulation of time/change. It always consists of three components, initial state $q_i$ (ice), intermediate state $q_\mathrm{inter}$ (ice with water) and final state $q_f$ (water). The line represents an abstract quantity, such as free energy.}\label{fig:F1a}
\end{figure}
%%%%%%%%%%%%%%%%%%%%%%%%%

When considering the description of change as a central axis, it becomes apparent that distinct classifications of change exist. Abstracting the notion of change reveals three fundamental phases: the {\it initial} state, the {\it intermediate} state, and the {\it final} state (See Fig.~\ref{fig:F1a}). For instance, in the phase transition of ice melting at room temperature, we can clearly identify the initial state as ice, the intermediate state as the melting process, and the final state as liquid water. Thermodynamics provides a rigorous theoretical framework for predicting such state transitions, but the specific elements included in the analysis lead to distinct theoretical formulations. Since the introduction of the second law of thermodynamics by Clausius, equilibrium thermodynamics has been established as a theory that exclusively deals with the relationship between initial and final states, while ignoring the intermediate state. This means that equilibrium thermodynamics can rigorously answer the question, "What happens \changed{after} ice is placed at room temperature?" but fails to address questions concerning "How does the ice melt?" or "When does the melting finish?". Theories that account for intermediate states require an entirely different framework, namely kinetics \cite{Landafshitz10,Kompaneyets} \footnote{Soviet physicists tends to call the non-equilibrium theory of statistical physics a term, {\it Physical Kinetics} \cite{Landafshitz10,Kompaneyets}.} and non-equilibrium thermodynamics \cite{Katchalsky, Kondepudi1998}. Crucially, these theories incorporate both entropy production and the concept of time to achieve a more comprehensive description of dynamical processes.

Notably, the two frameworks of equilibrium and non-equilibrium thermodynamics did not emerge simultaneously. The equilibrium thermodynamics was established in the mid-19th century \cite{Clasius1865}, whereas the non-equilibrium thermodynamics only emerged after 1931 \footnote{The statement that non-equilibrium thermodynamics began in 1931 refers to the fact that the explore for universally applicable laws governing all non-equilibrium states, although they were limited to the linear regime, began after the discovery of Onsager's reciprocal relations in 1931. However, the study of non-equilibrium phenomena themselves dates back much earlier.}. Prior to this, classical dynamics and equilibrium thermodynamics constituted the foundational theories, and the later development of non-equilibrium thermodynamics represented an effort to unify these two domains.

\begin{figure}[h]
\begin{center}
\includegraphics[width=12cm]{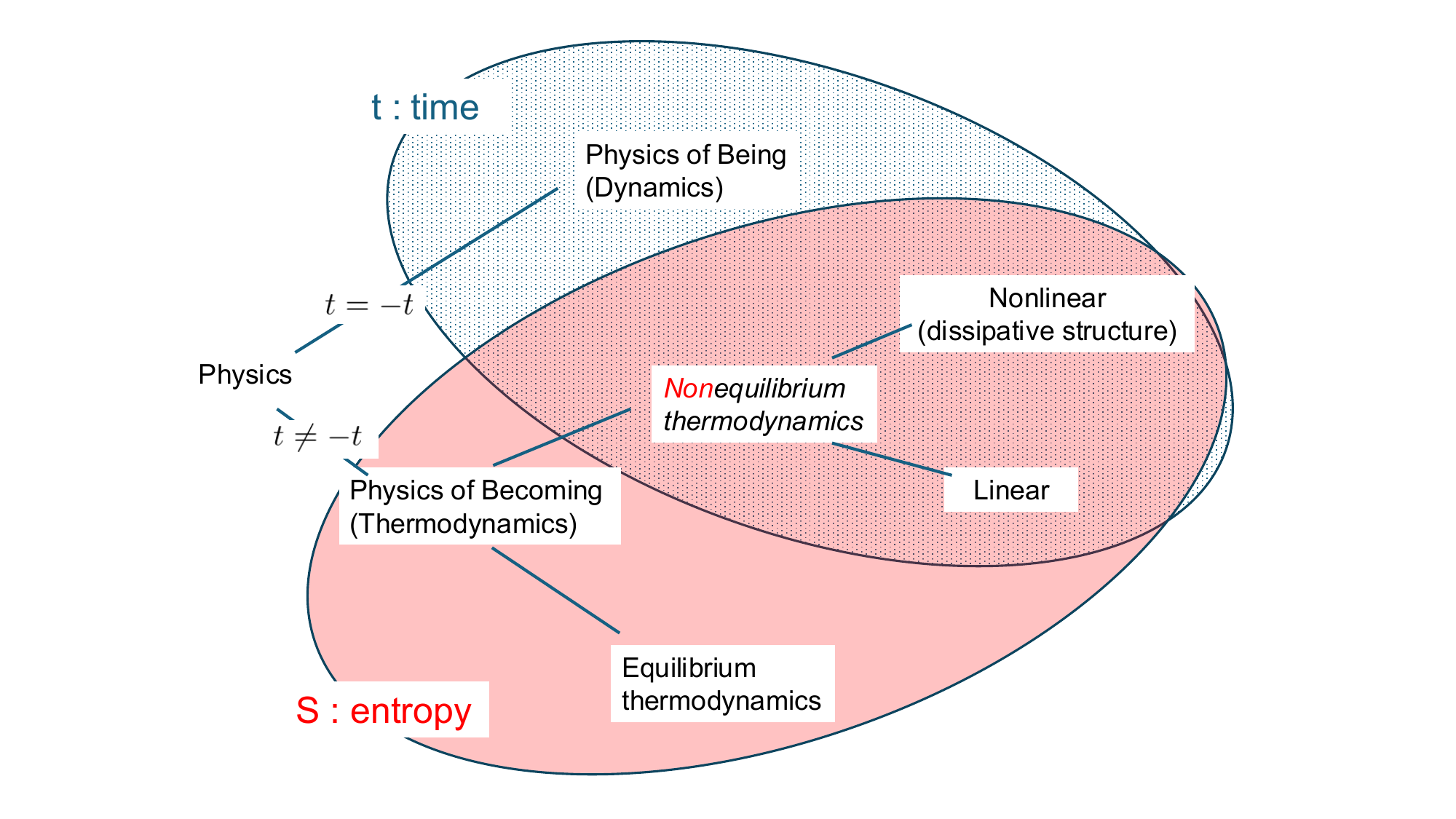}
\caption{The diagram of physics based on time-symmetry, equilibrium vs non-equilibrium, linearity vs nonlinearity. Physics, based on time symmetry, is divided into dynamics and thermodynamics. Thermodynamics is further split into equilibrium thermodynamics and non-equilibrium thermodynamics. Non-equilibrium thermodynamics, in turn, is further classified into linear and nonlinear regimes.  \label{fig:F1}}
\end{center}
\end{figure}

Examining non-equilibrium thermodynamics in the nonlinear regime presents an intriguing perspective. Non-equilibrium thermodynamics can be further categorized into two domains: linear non-equilibrium thermodynamics and nonlinear non-equilibrium thermodynamics (See Fig.~\ref{fig:F1}). The former applies to regions near equilibrium, where the assumption of linearity \changed{between the forces and flows} holds \cite{Kondepudi1998}. Although it deals with non-equilibrium systems, linear non-equilibrium thermodynamics operates under the principle of minimum entropy production, ensuring that the system evolves toward a state that minimizes entropy production, given the imposed boundary conditions. This implies the existence of a well-defined final state. However, in the nonlinear regime, such a final state may cease to exist under certain conditions. This phenomenon was first identified by Prigogine and colleagues in the Belousov–Zhabotinsky (BZ) reaction and was later demonstrated in a model known as the Brusselator. In the Brusselator model, an autocatalytic reaction pathway plays a crucial role. In conventional chemical reactions, reactants and products are molecularly distinct, as in the case of $A + B \rightarrow C + D$. However, in an autocatalytic reaction, the reactants and products include the same molecular species, as exemplified by $2X + Y \rightarrow 3X$. Mathematically, this results in a nonlinear term of second order in the reaction rate equation, forming a feedback loop where the reaction's outcome (final state) influences its cause (initial state). This distinguishes nonlinear systems from linear ones, where cause and effect are clearly separable. In nonlinear regimes, however, these elements become entangled. Such reaction pathways create conditions in which instabilities are amplified, leading to emergent chemical oscillations under appropriate conditions\footnote{In the linear regime, the linear relation of the phenomenological force and the flow, the minimum entropy production and the reciprocal relations are generally satisfied to get Lyapnov function while, in the nonlinear regime, these linearities are broken, then the Lyapunov function is not satisfied. See Ch.18, 19 of Ref. \cite{Kondepudi1998}}. When spatial diffusion processes are involved, self-organized pattern formation can occur, giving rise to structured spatial or spatiotemporal dynamics, which is {\it dissipative structure}.

%%%%%%%%%%%%%%%%%%%%%%%%%%%%%%%
%%%%%%%%%%%%%%%%%%%%%%%%%

\begin{figure}[h]
\includegraphics[width=12cm]{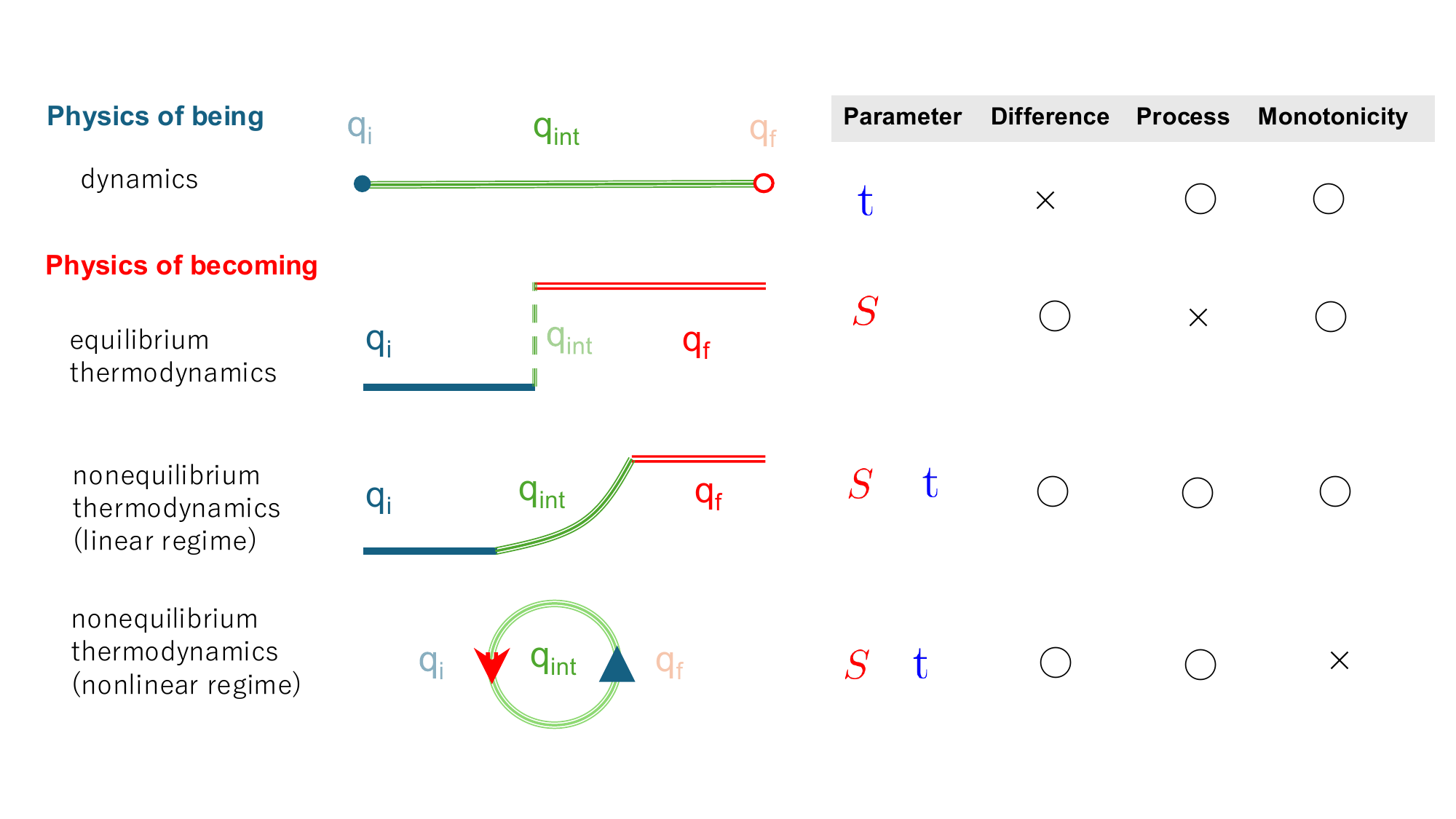}
\caption{The category of the concept of time in physics. The concept of time can be classified to four based on the combination of attribution of {\it difference}, {\it process} and {\it monotonicity}. Dynamics dealt with the time lacking in difference, while equilibrium thermodynamics deal with the time without process. Non-equilibrium thermodynamics in linear regime dealt with the time with all three attributes while non-equilibrium thermodynamics in nonlinear regime deal with the time without monotonicity. You will see that difference is mediated by entropy $S$, the process is mediated by time parameter $t$, and the monotinicity is mediated by linearity. }\label{fig:F2}
\end{figure}

By examining these theoretical frameworks, we can summarize the concept of time in physics as follows. First, change can be characterized by the existence of a {\it difference} that separates two states, such as the initial state and final state, or cause and effect, a {\it process} that continuously connects these two states, {\it monotonicity}, meaning that the process progresses unidirectionally from the beginning to the end. These three attributes---difference, process, and monotonicity---can be used to classify and characterize motion in physics (see Fig.~\ref{fig:F2}).

The concept of time in dynamics assumes the existence of a continuous {\it process} connecting the beginning and the end, exhibiting {\it monotonicity} from the initial to the final state. However, it lacks a discrete {\it difference} between the two states. This perspective is represented by the time parameter $t$, which underpins the notion of a reversible process. In contrast, within the physics of becoming, a {\it difference} exists between the beginning and the end, yet the evolution from the initial to the final state retains {\it monotonicity}. Equilibrium thermodynamics, however, lacks a concept of {\it process} that continuously connects these two states, with entropy playing an intermediary role. Thus, dynamics and equilibrium thermodynamics present two distinct temporal concepts: one where difference is absent and another where process is absent. Specifically, the process that connects the initial and final states continuously is governed by the time parameter $t$, while the difference between the two states is dictated by entropy $S$. It is only through the integration of these two aspects that the concept of time in non-equilibrium thermodynamics emerges. In the linear regime, the principle of minimum entropy production generally holds, ensuring that entropy production is minimized under given constraints. As a result, the system retains the three essential temporal characteristics: the difference between the initial and final states, the process as a form of continuity, and the monotonicity of evolution from the initial to the final state. However, in certain dynamics observed in nonlinear non-equilibrium thermodynamics, the system maintains the difference and the process but lacks {\it monotonicity}. As exemplified by autocatalytic reactions, the final state can influence the initial state, forming a feedback loop. This represents a temporal concept where the monotonicity between beginning and end is absent, resulting in an ongoing circular interaction where the two states, while qualitatively distinct, continuously generate each other. This circularity characterizes the notion of time in nonlinear non-equilibrium thermodynamics\footnote{It is important to note that nonlinearity does not always manifest in the form of autocatalytic reactions, such as those observed in the Brusselator model. For instance, in cases where nonlinear terms form similarity of the second kind \cite{Barenblatt2003}, pattern formation or propagating waves may not necessarily emerge. However, even in similarity of the second kind, the persistence of initial conditions and the divergence of correlations indicate a fundamental inseparability of multiple parameters. In this sense, they share common features. Ultimately, nonlinearity, in its mathematical abstraction, simply implies the absence of linearity, necessitating careful consideration of what specific type of nonlinearity is physically relevant.}.

While these classifications do not strictly separate different scientific disciplines, they allow for a major characterization of the concept of time in each fields. This characterization is based on whether the three essential properties---difference, process, and monotonicity---are present or absent. In other words, the concept of time in physics can be understood as an extended formulation of the aspects of time that we experience in daily life. A significant trend emerges from this analysis. In the early 20th century, the predominant physical theories were dynamics and equilibrium thermodynamics. Historically, non-equilibrium thermodynamics emerged later than equilibrium thermodynamics, and while the foundational theories of equilibrium thermodynamics have been well established, the fundamental theories of non-equilibrium thermodynamics remain in development. In particular, nonlinear non-equilibrium thermodynamics has yet to establish universal laws comparable to those in equilibrium thermodynamics and linear non-equilibrium thermodynamics. The asymmetry in the development of these fields arises from the characteristics of the problems intrinsic to each field and reflects how time is conceptualized.

Ultimately, physics is more adept at formulating well-defined time concepts that isolate specific aspects of time, while struggling with the comprehensive treatment of non-equilibrium time, which integrates all aspects. As seen in equilibrium thermodynamics, physics has demonstrated a particular strength in result-oriented formulations that disregard the process, further reinforcing its methodological inclination toward equilibrium-based approaches.
%%%%%%%%%%%%%%%%%%%%%%%%%%%%%%%
%%%%%%%%%%%%%%%%%%%%%%%%%

\section{The philosophy of dynamics and thermodynamics}

In the previous section, we examined the framework of time concepts in physics and found that the three fundamental elements---initial state, final state, and intermediate state (process)---can be classified based on how they are structured by the properties of difference, process, and monotonicity. In this sense, the temporal concepts of different fields within physics are constructed by the absence of certain attributes. This highlights the fact that the concept of time in physics is inherently divided and one-sided.

These characteristics of different theoretical frameworks, however, also reflect the underlying philosophical assumptions embedded within each field of physics. The reason Prigogine juxtaposed the physics of being (dynamics) with the physics of becoming (thermodynamics) was that he recognized that such theoretical classifications are deeply rooted in the intellectual traditions of physicists. This section focuses on the philosophical perspectives inherent in the physics of being (dynamics) and the physics of becoming (thermodynamics). Although these perspectives are often implicit rather than consciously acknowledged, as Prigogine pointed out, physicists tend to favor dynamics while positioning the physics of becoming lower in importance. This bias can be understood by examining how {\it dissipation} is positioned within each framework.

{\it Dissipation} plays a central role in the physics of becoming, namely in thermodynamics. Thermodynamics itself originated from the study of heat engines, where the goal was to convert heat into useful work, making it inherently engineering-oriented\footnote{It was said that Carnot's theory, considered to be the beginning of thermodynamics, was reportedly completely ignored by the leading physicists of his time \cite{Mendoza}. However, it was suggested that this neglect can be attributed to the fact that his approach to problem-setting differed significantly from that of mathematical physicists and was instead closer to engineering \cite{Yamamoto2009}.}. The primary concern in this field has been how efficiently heat can be converted into work, which ultimately led to the formulation of the concept of entropy. Within this framework, dissipation emerges as a quantity explained through entropy production. The second law of thermodynamics, which arose from the question of how entropy exchange (heat transfer from high to low temperatures) is compensated by entropy production (the conversion of work into heat), established entropy production as the fundamental cause of irreversible processes. However, entropy production itself appears in theoretical formulations as an epistemically "agnostic" quantity---a measure of unaccounted-for energy loss---rather than a physically concrete entity. The sources of such dissipation were often attributed to friction and other irreversible processes in heat engines, and these were traditionally regarded as engineering and technical imperfections\footnote{Typically it was seen in the modern textbooks such as Ref. \cite{Kittel,Atkins}.}. This historical perspective naturally led to the idea that dissipation is a non-essential imperfection that could ideally be eliminated through technological advancements.

In contrast, the physics of being, which is fundamentally based on reversible processes---namely, dynamics---does not regard dissipation as a central problem. Dynamics was originally developed from kinematics, which aimed to describe motion without considering its causes, focusing instead on its mathematical formulation. It was Newton who introduced causal forces, such as universal gravitation, into the study of motion. Following Newton, Euler, Lagrange, and other continental mathematicians refined and systematized the field \cite{Yamamoto1997}. The origins of dynamics are therefore deeply rooted in mathematics, and its fundamental equations naturally emerge from principles such as time homogeneity and spatial homogeneity. It focuses on the conservation of energy, treating work as a well-defined, quantifiable entity. In the fundamental equations of dynamics, no unknown quantity corresponding to dissipation exists. Consequently, these equations predict motion symmetrically in both the past and the future. Thus, within the fundamental laws of dynamics, there was never a logical necessity to incorporate dissipation into its foundational equations.

This absence of epistemic uncertainty at a fundamental level resonates with a worldview that assumes the universe to be static, deterministic, and mechanistic. Such a worldview envisions "all-embracing schemas, universal unifying frameworks, within everything that exists could be shown to be systematically- i.e.,logically and casually-interconnected, vast structures in which there should be no gaps left open for spontaneous, unattended developments, where everything that occurs should be, at least in principle, wholly explicable in terms of immutable general laws" \cite{Prigogine1984,Berlin}. This perspective implies that all natural phenomena are governed by underlying laws, which, once actualized, give rise to observable reality. Within such a worldview, the pursuit of natural laws resembles an archaeological endeavor---an attempt to uncover pre-existing rules that are already embedded in the fabric of reality. Events are seen as reversible and repeatable, reinforcing the idea that the world can be fully explored and exhaustively understood through these laws. The scientific worldview, in this sense, inherently aligns with the concept of reversible time, as it presupposes a static foundation of governing principles.

This is the worldview that dynamicists fundamentally assume. Given this historical trajectory and the relationship between humans and nature that this worldview presents, it is understandable why physicists, particularly those engaged in dynamical systems, find this framework intuitive. As a consequence, a significant conceptual gap emerges between dynamics and thermodynamics in their treatment of dissipation.

\begin{table}[t]
  \centering
  \begin{tabular}{ll}
    \hline
    Dynamics & Thermodynamics \vspace{.5mm}\\  
    \hline
     & 0th law : $T_A = T_B$ and $T_B = T_C$, then $T_A = T_C$ 
     \vspace{.5mm}\\
    1st law : $\vec{F}=0$ then $\vec{v}={\rm const.}$ & 1st law : $dU = dW +dQ$ \vspace{.5mm}\\
    2nd law : $\frac{d}{dt}\left(m \vec{v} \right) =\vec{F}$ & 2nd law : $d_\mathrm{i} S \geq 0$ \vspace{.5mm} \\
    3rd law : $\vec{F}_{21} =\vec{F}_{12}$  & 3rd law : $\lim_{T \to 0} S = 0$ \vspace{.5mm}\\  
    \hline
  \end{tabular}
  \caption{The comparison of fundamental laws of dynamics and thermodynamics. Note that dissipation, represented by $Q$ and $S$, is incorporated into the fundamental laws of thermodynamics.}
  \label{tb:t1}
\end{table}

Now let us compare the fundamental laws of dynamics and thermodynamics in the Table \ref{tb:t1}. The fundamental laws of dynamics, such as Newton's three laws \cite{Haar} and the Lagrange equations, are derived from time symmetry and spatial symmetry. However, the concept of dissipation is not incorporated within these fundamental laws. Instead, friction and air resistance are introduced phenomenologically as additional terms to adjust theoretical models to match real-world observations. In other words, dissipation in dynamics is merely {\it optional}---it is not regarded as a fundamental concept that forms the core of dynamical theory.

In contrast, thermodynamics treats dissipation as a crucial concept that is explicitly incorporated into its fundamental equations, most notably through the second law of thermodynamics. The first law represents the principle of energy conservation, while the second law establishes the directionality of natural processes and asserts the existence of irreversibility. In this sense, thermodynamics can be regarded as the science of dissipation. Thus, dissipation occupies fundamentally different positions in the two major fields of physics. It is important to note that dissipation is indeed considered in dynamics. However, the crucial distinction is that, in dynamics, dissipation is treated as a convenient approximation---a secondary modification added to idealized equations to align with empirical reality. At its core, it is viewed as a redundant or extraneous concept rather than a fundamental principle.

This fundamental gap between the laws of dynamics and thermodynamics leads to an interpretational issue when thermodynamics is analyzed from the perspective of dynamical theory. From this standpoint, a specific interpretation of the second law of thermodynamics emerges, often referred to as the {\it subjective interpretation of the second law}.

\begin{quote}
Physicists have generally regarded the dynamical description as fundamental, considering the Second Law to be a consequence of a certain approximation applied to dynamics. [$\cdots$] For example, Max Born explicitly stated that "Irreversibility is therefore a consequence of the explicit introduction of 'ignorance' into the (dynamical) fundamental laws"\cite{Born}. (From Being to Becoming, Japanese edition, Ch.10, p. 220 \cite{Prigogine1980b})
\end{quote}

What is clearly reflected here is the attitude that the second law of thermodynamics, despite being a fundamental principle in thermodynamics, does not hold an essential status. From the perspective of dynamical theory, which fundamentally assumes reversibility, the second law is regarded as a non-essential consequence, emerging only when human intervention is introduced. From this viewpoint, dissipation is considered a special case, and its existence does not in any way challenge the foundational worldview of dynamics. This perspective also extends to the belief in the complete predictability and accessibility of the world.

As formulated in thermodynamics, dissipation is the entropy inherently generated within a system when the conversion of heat into work fails to achieve maximum efficiency. Given this origin, dissipation is indeed an approximate concept, positioned fundamentally as a constraint within physical systems. This explains why dynamicists tend to disregard dissipation, and their dismissive attitude toward it reflects a similar stance toward the limits of their worldview.

For dynamicists, limitations in their theories are not fundamental limitations. Instead, such constraints are seen as practical and technical challenges that are ultimately resolvable through advancements in knowledge and technology. These constraints do not impose any fundamental restrictions on the dynamicists' worldview itself. In other words, they are real-world, technical limitations, but they do not imply any inherent limitation of the dynamicists' perspective on reality.

\section{Naive agnostic attitude and essential agnostic attitude}

\changed{In the preceding section, we examined the difference in worldview between dynamics and thermodynamics. While this difference stems from the foundational equations that define each theory, the construction of these equations itself reflects not only the physical laws they are based on, but also the worldview underpinning those laws, as well as the epistemological stance of the physicists who share that worldview. Although such perspectives on the nature of physical reality do not constitute practical distinctions and are often regarded merely as matters of interpretation---hence receiving limited attention in the hierarchy of physics---they nevertheless warrant scrutiny in the present context. This is because the issues raised by Prigogine concern not only physics itself, but also the attitudes of physicists toward their discipline.}

\changed{In order to clarify this point, I aim to formulate the position of dynamicist in terms of an agnostic stance. Most scientific theories, more or less, possess intrinsic limitations defined by their domains of applicability. Within the natural sciences, such domains of applicability are ultimately grounded in mathematics and empirical observation. This attitude of avoiding asserting the validity of claims in the absence of evidence derived from certain methods exemplifies an agnostic attitude in the sense advocated by Thomas Huxley.}

\begin{quote}
\changed{This principle [of Agnosticism] may be stated in various ways, but they all amount to this: that it is wrong for a man to say that he is certain of the objective truth of any proposition unless he can produce evidence which logically justifies that certainty. This is what Agnosticism asserts; and, in my opinion, it is all that is essential to Agnosticism. (Agnosticism and Christianity,
Collected Essays V p. 12 \cite{Huxley1899}).}
\end{quote}

\changed{From Huxley's formulation, one can discern an agnostic stance as a defining feature of the scientific attitude---namely, an epistemic modesty that refrains from making definitive claims about matters not grounded in mathematical reasoning or empirical observation. In this respect, physics---being grounded in theories that rely on experimental observation---necessarily entails intrinsic limits on the scope within which its claims can be considered valid. This suggests the existence of an intrinsic {\it attitude} or {\it strategy} toward its inherent limitations. As we have seen in the previous section, the way in which these limits are positioned differs between dynamics and thermodynamics.}

When formally redefining the attitude held by dynamicists, it can be described as a perspective in which the world is objectifiable, where repeatability is always guaranteed, and where everything can be exhaustively understood in principle. However, what is even more critical is that the very existence of this idealized static world serves a significant function within this framework. Dynamics inherently acknowledges the gap between the ideal and reality within its worldview. Dynamicists are, of course, aware of the fact that the real world differs from the ideal, and that dissipation exists as an undeniable phenomenon. However, {\it what is crucial is not the existence of the gap itself, but rather how this gap can be systematically related to their theoretical framework}. In this sense, dynamics possesses the means to systematically push aside the discrepancy between ideal and reality by attributing it to technical constraints. The subjective interpretation of the second law functions as a rhetorical strategy that reduces dissipation to a non-ideal state of dynamics, merely a result of technical limitations. In other words, dynamics manages to objectify even its own limitations, thereby ensuring that no fundamental constraint ever extends to the worldview itself. The more unconscious the belief in a static, deterministic world becomes, the more any factor that obstructs this worldview is trivialized. The way dynamics treats dissipation is a direct manifestation of this mindset. This perspective culminates most prominently in the subjective interpretation of the second law, where dissipation, which fundamentally challenges the static nature of the world, is classified as a mere technical constraint and thus non-essential. This attitude can be formally described as a {\it naïve agnostic} stance or attitude, where any fundamental challenge to the dynamicists' worldview is systematically dismissed as a solvable technical issue rather than an inherent conceptual limitation.

It was Prigogine who sharply reacted to the unconscious assumptions held by dynamicists. From the very beginning, Prigogine did not merely critique physics as a discipline but also regarded the physicists themselves as part of the problem. His criticism of how dissipation is unjustly relegated to a low status within dynamical theory stems from his recognition, during the development of non-equilibrium thermodynamics, of the constructive role of dissipation.

In equilibrium thermodynamics, dissipation had previously been understood only negatively, as a phenomenon that reduces usable work due to friction or diffusion, representing uncontrollable energy loss and a transition from order to disorder. However, in non-equilibrium thermodynamics, particularly in nonlinear regimes, dissipation was found to contribute to the formation of order through instability. This paradoxical idea---that dissipation, which had been viewed purely as destructive within equilibrium theory, could have a constructive role in non-equilibrium systems---became central to his concept of dissipative structures.

\begin{quote}
 In far-from-equilibrium conditions we may have transformation from disorder, from thermal chaos into order. [$\cdots$] We have called these new structure {\it dissipative structure} to emphasize the constructive role of dissipative processes in their formation. (Order Out of Chaos, p. 12 \cite{Prigogine1984})
\end{quote}

As Prigogine continued developing dissipative structure theory as a model for biological systems, his dissatisfaction with dynamical theory, which treated dissipation as merely a secondary phenomenon, naturally deepened. The marginalization of dissipation in dynamics is, in essence, a marginalization of the very origins of life itself. Just as dynamical theory reduces dissipation to an insignificant byproduct, it similarly reduces life itself to a mere incidental phenomenon, treating friction and other irreversible processes as non-essential.

\begin{quote}
    According to the subjective interpretation, chemical affinity, heat conduction, viscosity, all the properties connected with irreversible entropy production would depend on the observer. Moreover, the extent to which phenomena of organization originating in irreversibility play a role in biology makes it impossible to consider them as simple illusions due to our ignorance. Are we ourselves-living creature capable of observing and manipulating-mere fictions produced by our imperfect senses? Is the distiction between life and death an illusion? (Order out of Chaos, p.252 \cite{Prigogine1984}.)
\end{quote}

This leads to two possible perspectives. The first, represented by dynamicists, maintains that dissipation is merely a peripheral phenomenon and accepts a mechanistic interpretation of life---reducing it to a system that can be fully explained through deterministic dynamics. The second, in contrast, recognizes dissipation as a fundamental principle akin to the laws of thermodynamics and seeks to integrate dynamics and thermodynamics into a unified framework. This latter perspective is precisely Prigogine's stance, which served as the primary motivation for his third phase of research.

\begin{quote}
    We do not seek to explain the second law as a mere approximation introduced into dynamics or as an apparent fact arising from our ‘ignorance'. Instead, we take the second law as a {\it fundamental physical fact} and explore the transformations that this assumption brings to our concepts of spacetime and dynamics.
    (From Being to Becoming, Japanese edition, Ch. 10, p. 221 \cite{Prigogine1980b}.)
\end{quote}

Prigogine's motivation for elevating the second law of thermodynamics to the level of a fundamental physical law stems from a deeper challenge to the worldview underlying dynamics itself. He saw physics as an inseparable interplay between its theoretical framework, its practice, and the worldview held by physicists. In this sense, Prigogine's work---especially in Order Out of Chaos---inevitably carried not only scientific but also cultural and revolutionary implications regarding how humans understand nature. However, this philosophical critique of physics, which could also be seen as an attempt to enlighten physicists, was unlikely to resonate with dynamicists who held a naïve agnostic stance---one that fundamentally separates physics from the physicists practicing it. For them, Prigogine's approach might have appeared grandiose or even misplaced. Yet, when considering the broader historical and conceptual context of his ideas, it becomes clear that Prigogine's arguments possess a consistent and compelling logic.

At its core, Prigogine's stance can be formally defined as an acknowledgment of the objective existence of irreversibility in nature---an acceptance that certain processes in nature are fundamentally beyond human control. This also implies an acceptance of the principle that nature itself is something that humans can never fully exhaust or encapsulate in a finite framework. Furthermore, it affirms that nature is not something that humans can fundamentally dominate or control, not merely on a technical level but at an ontological level. In this sense, his view does not simply concern physics but extends to the fundamental limitations of human knowledge itself. This perspective, in contrast to naïve agnosticism, can be formally defined as {\it essential agnosticism}.

In a naïve agnostic stance, as seen in the subjective interpretation of the second law, dissipation is treated as a special exception, and the constraints imposed by nature on human knowledge are assumed to be non-essential---at most, practical limitations that could theoretically be eliminated. It is crucial to note that for dynamicists who adhere to this perspective, the limits of knowledge are always technical rather than fundamental. This means that even if their idealized theories fail to describe reality, those failures are treated as mere anomalies, never as a challenge to the fundamental validity of their worldview. In this framework, scientists believe that they can continue to analyze, control, and understand the universe from a purely objective standpoint, akin to a \changed{God's-mind} perspective that remains unshaken by any empirical failures. If they can successfully trivialize the nature of human limitations, then those limitations can be dismissed as non-essential.

Essential agnosticism, on the other hand, directly challenges this idealization at a fundamental level. It rejects the assumption that humans possess an absolute, external viewpoint from which to objectify the universe. Instead, it proposes that humans exist in an intrinsic, reciprocal relationship with nature, in which we are not external observers but rather players in a game, constrained by the very rules that govern it.

Crucially, Prigogine does not view these fundamental limitations as negative or restrictive. Rather, he argues that these constraints are what make life possible in the first place, and that they are an indispensable part of human existence. In contrast, the naïve agnostic stance trivializes and dismisses these constraints precisely because it can only perceive them as obstacles rather than as essential features of reality. 

\section{Spatialization of physics}

In the preceding section, we formulated the attitude of dynamicist—particularly those who adopt a subjective interpretation of the second law of thermodynamics---as a naïve agnostic attitude, and contrasted it with Prigogine's position, which we designated as an essential agnostic attitude. We will see that this different attitudes give rise to a fundamentally different way on physics. The former is spatialization, the latter is temporalization. Prigogine himself once suggested that his ambition was to temporalize physics:

\begin{quote}
I would like to emphasize that I am not at all speaking about the end of science or the end of time. Physicists like Hawking are trying to come up with an ultimate unified theory with which we will be able to explain everything. In a sense they are trying to understand the mind of God. I consider this a great naivete.

Physicists have long been averse to thinking about the problem of time because it was believed that the absence of time would itself be the most compelling evidence that we had begun to approach the mind of God. This is the position of Einstein and Hawking. Hawking has inherited Einstein's vision and is trying to make physics into a kind of geometry, to spatialize it. I, on the other hand, am trying to temporalize physics. (Time and Creation : An Interview with Iliya Prigogine \cite{Prigogine1998}) 

\end{quote}

Prigogine did not provide a precise formulation of what he meant by the temporalization of physics. Nevertheless, we may attempt to interpret this concept in light of his philosophical attitude. In this section, we focus on the spatialization of physics, drawing on the insights of Henri Bergson, whose ideas were expected to have significantly influenced Prigogine's thinking.

The motivation underlying Prigogine's intellectual stance can, in part, be attributed to the insights he developed through his work on non-equilibrium thermodynamics---particularly his recognition of the constructive role of dissipation. However, it is also grounded in the insights that have derived from his philosophical interests \cite{Kitahara,Prigogine1937a,Prigogine1937b,Prigogine1937c,Spire}. From his earliest publications, Prigogine contrasted the universality of physical laws with the dynamic evolution observed in biological systems and human societies. It seems that this is derived from his philosophical view originated from before he became a scientist, particularly from Bergson, as it is explicitly referred in his Nobel lecture\cite{Nobel}.

Bergson's philosophy revolves around the concept of time, and at its core, it contains a critique of how time has been traditionally conceptualized---not only by philosophers but also by natural scientists. His argument is that most theories of time, whether in philosophy or science, fail to adequately capture the true nature of time.

\begin{quote}
    Not one of them [philosophers] has sought positive attributes in time. They treat succession as a co-existence which has failed to be achieved, and duration as a non-eternity. (The Creative Mind: An Introduction to Metaphysics, p.10 \cite{Bergson1934}). 
\end{quote}

\changed{Bergson's critique---that philosophers have failed to attribute positive to time---finds a striking parallel in Prigogine's assertion that dynamicists have similarly neglected to assign any constructive role to dissipation.}

\changed{To clarify this point, Bergson introduces the concept of dur\'{e}e---translated as duration---as the nature of time. Duration emphasizes time not as a series of discrete, measurable intervals, but as a flowing continuity and as a process unfolding in real time. In this view, the essence of time lies in its incompleteness and openness, which is something in the {\it process of happening}. Time as duration, in Bergson's philosophy, is not a sequence of positions in space but a real and continuous becoming, an unfinished process that constitutes the very nature of time itself.}

\changed{Various attempts have since been made to formally express such a temporal conception in terms of the grammatical aspect, where incompletion is considered intrinsic to the structure of time \cite{sugiyama2006,Hirai2022,Hirai2024,Lapoujade,During}. Yet, according to Bergson, humans are inherently constrained to conceive of time only as something already completed or objectified. This limitation---our tendency to derscribe time as a finished, static entity---is what Bergson denounces as the {\it spatialization of time}. Once time is objectified and treated as an object of scientific inquiry, it is inevitably stripped of its dynamic, flowing character and replaced with a spatial, measurable substitute. This is the core of Bergson's accuse about the spatialization of time.}

\changed{On the other hand, Bergson articulates the {\it incompleteness} inherent in time with a characteristic formulation \cite{Bergson1922}: we often say, "The child becomes a man," but in truth, he argues, "There is becoming from the child to the man." What is essential in this distinction is Bergson's conception of {\it process}, namely becoming (devenir), which is not defined by two independent points---a beginning and an end---connected by an interval. Rather, what exists first is the {\it process of becoming itself} (=becoming from the child to the man), a co-presence of beginning (the child) and end (the man), from which the notions of "beginning" and "end" derive as actualization. This {\it heterogeneity} is essential on the concept of duration. On the other hand, {\it the spaitialization of time is nothing but the elimination of this heterogeneity.} In this view, the "intermediate" or "in-between" is not a static element but a dynamic substance that can only exist as the process itself---a duration within which difference unfolds \footnote{This notion of process is further developed by Gilles Deleuze, who formalizes Bergsonian duration as a principle of differentiation. Deleuze writes that "what makes difference is no longer that which differs from something else, but that which differs from itself" \cite{Deleuze1956}. In this sense, the process is not constituted by external contrast between two distinct entities but by the process of differentiation itself.}.}

\changed{This insight is suggestive and  particularly consistent with the analysis of time in physics presented in Figure~\ref{fig:F2}. As emphasized there, the foundational frameworks of physics have been established either within equilibrium thermodynamics, which lacks the element of {\it process}, or within dynamics, which lacks the element of {\it difference}. A distinguishing feature of both conceptions of time is the absence of the {\it heterogeneity} of opposing elements. It is suggestive that these two domains, dynamics and equilibrium thermodynamics, comparatively succeeded in their foundation in comparison with the others in which the heterogeneity is essentially involved. It suggests that the successful physical domain is that are well spatialized in the way Bergson fromulated. Non-equilibrium theory can thus be interpreted as an attempt to reintegrate into physics the notion of heterogeneous process.}

%\changed{Non-equilibrium theory can thus be interpreted as an attempt to reintegrate into physics the notion of heterogeneous process understood as a state of coexistence---a dimension that had been excluded in earlier formulations. However, this attempt faces inherent difficulties, owing to what might be described as the structural tendency of physics to spatialize time. This inclination is not merely historical or contingent but reflects a deep-seated epistemological orientation within the physical sciences. Although there is little evidence that Prigogine explicitly identified this issue in terms drawn from Bergson, the trajectory of his work suggests a clear resistance to the spatialization of time. His efforts to extend the conceptual foundations of physics through non-equilibrium processes can be seen as moving in a direction that implicitly counters the very reduction of time that Bergson had critiqued.}

\section{Temporalization of physics}

\changed{Prigogine's critique of the spatialization of time in physics is directed, in its most immediate form, at the traditional underestimation of dissipation, the subjective interpretation of the second law. However, Prigogine's criticism ultimately extends beyond this conceptual hierarchy to encompass the deeper human attitude that naturalizes such a worldview---an attitude that regards nature as ultimately knowable, closed, and static. I have characterized this orientation as a naïve agnostic stance, in which the limits of human understanding are seen merely as technical constraints rather than as indicators of a more fundamental epistemological limit. In contrast, I identify Prigogine's position as an essential agnostic stance, which acknowledges the intrinsic uncertainty, openness, and temporal irreversibility of natural processes. This stance, I argue, represents a call for the restoration of the essential nature of time within physics. Prigogine himself seemingly referred to this effort as the {\it temporalization of physics}, a phrase he once hinted at in an interview \cite{Prigogine1998}. While he did not offer a formal elaboration of this notion, we may nonetheless explore the concept in a broader context, with the aid of the philosophy of Bergson and others. At this point, the subject appears to play an essential role. Once he said:}

%\changed{From these, we can understand the notion of an open world as corresponding to what I have termed the essential agnostic stance---a posture that accepts the principle of irreducible unknowability and the inexhaustibility of nature. In contrast, the idea of a closed world implies the absence of such fundamental ignorance; it corresponds to the naïve agnostic stance, which presumes that the world is, in principle, fully knowable and exhaustible. This contrast encapsulates two opposing epistemological attitudes toward the nature of reality: one that embraces openness and indeterminacy, and another that seeks closure and complete intelligibility.}

\begin{quote}
    
Demonstrations of impossibility, whether in relativity, quantum mechanics, or thermodynamics, have shown us that nature cannot be described "from the outside," as if a spectator. (Order Out of Chaos, p. 300 \cite{Prigogine1984}.)
    
\end{quote}

\changed{Particularly from this, the role of the subject is implicitly emphasized. The notion of an {\it open} world ultimately refers to the attitude of the human subject who stands in relation to the world. The role of the subject thus emerges as central to the very constitution of time. In the context of physics---where the aim is to render nature knowable and predictable---how the subject is positioned becomes a question of foundational importance. Conventionally, the subject is either excluded or concealed in physics. This exclusion is not accidental: it reflects the underlying methods of physics to objectify the nature. To make nature an object of scientific inquiry, the subject must be isolated from the nature. For instance, Davis points out that discussions on the direction of time and the arrow of time often arise because they conflate the vague psychological notion of "becoming" with the objective and rational physical concept of time asymmetry \cite{Davis}.}

\changed{However, both Bergson and Deleuze---who developed his own philosophical project through a distinctive interpretation of Bergson---recognized that the subject plays an essential role in the constitution of time. Deleuze, in particular, focuses on Bergson's concept of virtuality (la virtualit\'{e}), which it can be formulated as a degree or intensity of the unfolding process, or the degree of the process in which "something possessing the virtue or the power to be actualized, has not been actualized" according to Kimura \cite{Kimura1996} \footnote{Kimura discussed the notions of actuality and reality, focusing on the concept of virtuality. He said, "'Reality' is something that is objectively objectified through public recognition and, as a shared norm of a community, places certain constraints on the actions and judgments of its members. [$\cdots$] In contrast, 'actuality' belongs entirely to the practical act of 'living', {\it actio}, which is the individual, direct endeavor of life. [$\cdots$] While reality is expressed in the 'past tense' or 'perfect tense', actuality unfolds only in the 'present tense'---or, more appropriately, the 'present imperfect' in English. [$\cdots$] If reality is an indicator of beings, then actuality is a characteristic of becoming itself, and cannot be a marker of existence in any form. [$\cdots$] The state in which actuality has not yet been realized as actuality is 'virtuality'." \cite{Kimura1996}}. It is through this degree---heterogeneity of the process of becoming---that Deleuze locates the separation between subject and object. In other words, the subject and object are not pre-given, discrete entities, but emerge through the actualization of the process.}

\begin{quote}
Bergson means that the objective is that which has no virtuality - whether realized or not,  whether possible or real, everything is actual in the objective. [$\cdots$] In other words, the subjective, or duration, is the virtual. To be more precise, it is the virtual so far as it is actualized, in the course of being actualized, it is inseperable from the movement of its actualization. ( {\it Bergsonism} pp.41-42. \cite{Deleuze1966})
\end{quote}

\changed{Deleuze and Bergson conceive the distinction between subject and object as fundamentally virtual---that is, as a difference emerging along a continuum of actualization. The subject is characterized as being in the process of becoming, retaining its virtuality, whereas the object is understood as something that has already lost this virtuality, having become fully actualized. In this sense, the distinction between subject and object is not absolute, but rather defined by degrees of the intensity of virtuality. Consequently, both subject and object can be understood as two polar separated by the continuous process of the actualization. This interpretation has also been explored by Kimura \cite{Kimura1996}, who identified this kind of actualization as central to the framework of his own "medizinischen Anthropologie" (phenomenological–anthropological psychopathology \cite{Fukao, Phillips, Kimura, Aida}), which resonates with the work of Viktor von Weizs\"{a}cker, who stress the introduction of subjectivity in his theory of life \cite{Weizsacker, Weizsacker1946}.}

\changed{The subject immersed within the process of an event plays an essential role in the conception of time as a flow---as something in the process of occurring. This formulation implies that such an occurrence-in-progress cannot be meaningfully described without reference to a subject who experiences it. In other words, the notion of something "happening" presupposes a vantage point, which is subject, from which the unfolding is lived and experienced. It is here that the inseparability of temporal becoming and subjectivity becomes apparent.}

Considering this in physics, I believe that one example of this phenomenon in which the process of becoming is apparent is the pitch drop experiment \cite{Parnell}. Pitch is a general term for substances with extremely high viscosity, and the experiment involves pouring pitch into a funnel and observing its flow over time. Initially, even after being placed in the funnel, the pitch shows no visible signs of flow. Over the course of days, months, or even years, it appears to behave entirely as a solid, showing no fluid-like characteristics. However, after 75 years of continuous observation, seven drops of pitch were recorded. The estimated viscosity of pitch was $2.3 \times 10^8~{\rm Pa~s}$, which was 230 billion times larger than the viscosity of water. This observed flow behavior is evidence of its fluid-like nature.

What makes this experiment particularly intriguing is that the supposedly objective classification of a substance as either a solid or a fluid appears to change depending on the subjective scale of observation time. Ultimately, this phenomenon can be understood through the {\it intermediate asymptotics} \cite{Barenblatt2003, Barenblatt1972, Goldenfeld1989, Goldenfeld1992,Barenblatt1996, Barenblatt2014,Maruoka2023a} of the Deborah number \cite{Reiner}, ${\rm De} = \frac{\tau}{T}$, a dimensionless number defined as the ratio between relaxation time $\tau$ and observation time $T$ \cite{Maruoka2023}. Supposing the volume of the flowed pitch, $V$, the flow is described as $\frac{d V}{d T}$. When ${\rm De} \gg 1$, the relaxation time is much longer than the observation time, meaning that no fluid-like behavior is observed, and the material appears solid as $\frac{d V}{dt} = 0$. When ${\rm De} \sim 1$, the relaxation time and the observation time are of the same order, allowing fluid-like behavior to emerge as $\frac{d V}{dt} > 0$, which is summarized as follows \footnote{The two asymptotic expression can be described by multiple methods \cite{Holmes}, as shown in Ref. \cite{Maruoka2023}.},
\begin{eqnarray}
    \frac{dV}{dT} = \begin{cases}
     0 ~\left({\rm De} \gg 1 \right)\\
     \frac{\pi d^4 \rho g}{128 \mu}\left(1+\frac{h}{l} \right)~\left({\rm De} \sim 1 \right). 
    \end{cases}
\end{eqnarray}

However, this does not mean that the classification of pitch as either a solid or a liquid is purely subjective. The determining factor ultimately involves the relaxation time of the material itself, meaning that the physical properties of the substance must also be taken into account. What the pitch drop experiment fundamentally reveals is that the properties of a material are determined by the scale at which the subject (observer) and the object (material) interact. In this sense, the observer is inevitably {\it complicit} in defining the attributes of pitch. This {\it complicit} \footnote{The term {\it complicit} here is used to highlight the {\it inevitable} and {\it inseparable} contribution of the subject in determining the properties of materials.} relation between observer and material is the fundamental insight offered by the pitch drop experiment. 

For instance, glass has a relaxation time of 19,000 years \cite{Welch}. Following the same logic, glass would exhibit fluid-like behavior on a 10,000-year time scale. However, for humans, glass is perceived as a solid, because in our daily experience, glass does not flow. The reason for this is that the human interaction time with glass typically ranges from a few hours to at most a lifetime of several decades. Within this human time scale, glass has not fully manifested its potential behaviors, yet what is meaningful for humans is the glass as it appears within the constraints of human finiteness---as something in the midst of a process that is involving us. In this sense, humans play a role in determining the perceived properties of glass. \footnote{Glass has long been a subject of research within the physics community and remains a topic of ongoing debate. The present formulation is grounded in the concept of the Deborah number and the framework of intermediate asymptotics.}.

\changed{What must be emphasized here is that this is not merely a description of a transition from state A to state B. Rather, the transition itself---from A to B---constitutes the very reality of glass. It is precisely the heterogeneity of these two states that gives rise to the property we recognize as "glass." In this view, glass is a {\it temporal} material in the sense that it is retaining its heterogeneity. It is not a fixed object but a material essentially situated within a process. It acquires its identity as "glass" for the human subject precisely because it exists in a state of becoming in the scale range of the human. Crucially, this process of becoming only acquires ontological contribution through the presence of a subject who experiences it. In this sense, the attributes of the object---here, glass---are made possible only through a {\it complicit} contribution with the subject.} Here we can find that there is heterogeneity between two states (solid or fluid) derived from the degree of non-equilibrium, and heterogeneity between the subject and the object derived from the degree of complicity. These two heterogeneities generate an actuality of the phenomenon. As far as the heterogeneity is maintained, the subject/phenomenon is forced to be {\it localized}.

By contrast, in a state of thermodynamic equilibrium, the process has reached its terminus. At all relevant scales, the object ceases to undergo change, and the subject's role in constituting its identity effectively comes to an end. Here the heterogeneity has already relaxed. The object, now fully actualized, no longer requires the roll of the subject explicitly. 

Note how the heterogeneity observed in the pitch drop experiment is consistent with the concepts of Bergson and Deleuze. Scale-dependent phenomena, such as the pitch drop experiment, reveal the role of heterogeneity and subjectivity in physics. Thus, we identify two forms of subjectivity: one should be the {\it spatialized subject}, and the other should be the {\it temporal subject}. These two subjects differ fundamentally in their degree of heterogeneity.

\begin{figure}[h]
\includegraphics[width=12cm]{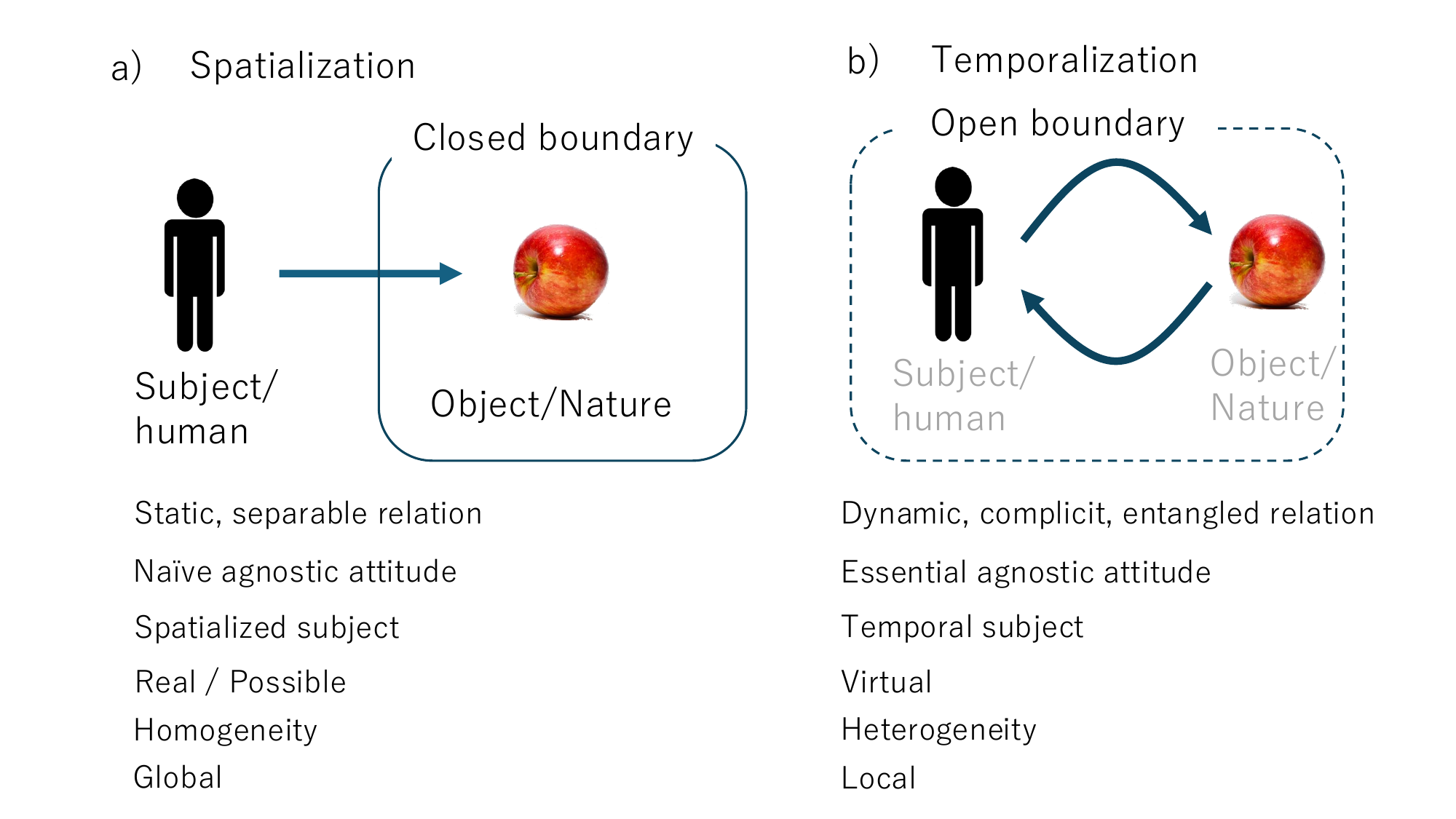}
\caption{A comparison between spatialization a) and temporalization b), accompanied by schematic figures illustrating the relationship between subject/human and object/nature, along with key terms and concepts. Spatialization is static and closed in the sense that all the process of becoming has been realized to eliminate the heterogeneity of subject and object, where the virtuality has been lost. In contrast, in temporalization, these two elements are entangled (i.e., heterogeneity) as a unified event whose boundary remains open. This complicit, entangled relation to the event is represented by a circle.}\label{fig:F3}
\end{figure}

Now let us summarize what the spatialization and the temporalization are. The state of "\changed{becoming}" of non-equilibrium refers to a situation in which the subject and object are {\it inseparably entangled} in an unified event \footnote{Weizs\"{a}cker formulated the connection maintained between the subject and the environment as 'Koh\"{a}renz', in which movement and perception are {\it entangled} and taken as {\it one} act. Such an 'entanglement' (Verschr\"{a}nkung) was occasionally illustrated by a {\it 'circle'} (Kreis) in his Gestaltkreis theory \cite{Weizsacker}. }, the boundary between the subject and the object is obscure, and {\it open} depending on the degree of complicity. This corresponds to the subject in an essential agnostic attitude. This stands in stark contrast to the dynamicists' perspective, which is naive agnostic attitudes assuming a clear separation between humans and observed phenomena where the boundary is sharpened to be {\it closed}. All these contrasts are summarized in Fig. \ref{fig:F3}. 

%This idea can also be extended to the relationship between humans and nature, as suggested by the pitch drop experiment. In this experiment, the boundary between the observer and the phenomenon is not sharply defined, revealing a cohesion between subject and object (See Fig.~\ref{fig:F3}).

%Instead, in non-equilibrium systems, humans and events are partially unified. This perspective aligns with Prigogine's repeated emphasis on a new scientific outlook, one in which scientists themselves are embedded in the very world they seek to describe.

%Prigogine recognized that the development of relativity and quantum mechanics exposed fundamental limitations that highlight the {\it entanglement} of subject and world. He regarded the Second Law of Thermodynamics as another such limitation---a constructive constraint on our understanding of the universe. This perspective fundamentally transforms the role of physicists: no longer mere external observers, they are now participants within the system. Whether in relativity or quantum mechanics, these constraints on the physical world imply a profound complicit relation between subject and reality. This suggests, at its core, that the world is not a closed system but an open one. As a result, humans and nature form a circular, reciprocal relationship, reinforcing the inherent openness of the universe.

%%%%%%%%%%%%%%%%%%%%%%%%%%%%%%%
%%%%%%%%%%%%%%%%%%%%%%%%%

Taking these insights into account, we can redefine Prigogine's vision of time in physics. It represents a shift from a physics centered on staticity to a physics centered on dynamism. This transition also restores the rightful significance of previously marginalized concepts such as dissipation and uncertainty, which had been treated as purely negative within the static framework of classical physics. Moreover, this shift inevitably transforms the relationship between humans and nature. In a world fundamentally governed by non-equilibrium, humans can no longer perceive themselves as separate from the world. Engaging with dynamism reveals the blurring of conceptual boundaries, and consequently, it brings to light the fundamental interconnectedness between humans and nature\footnote{Although this emphasis on the relationship between subject and nature is somewhat less pronounced in The End of Certainty \cite{Prigogine1997}.}. Ultimately, the fact that we are {\it local} beings entangled with this world serves as a direct demonstration that we can never fully encapsulate and exhaustively comprehend the entirety of reality. This constitutes the epistemological positioning of unknowability within the framework of essential agnosticism. Physics is temporalized in the sense that scientists come to recognize the nature of time as the actualization of time---that is, time that preserves its inherent heterogeneity. 

%\footnote{These views are quite resonate with the antholopolgy of Ende \cite{Ende}, and the philosophy of Ushiki \cite{Ushiki,Maruoka2017}.}.} 

The naïve agnostic attitude and the essential agnostic attitude thus represent two distinct trajectories for the development of physics. The former seeks to maintain the objectivity of physics by relegating its empirical ignorance to the realm of subjectivity, thereby reinforcing an ostensibly secure and objective foundation. Now we see that the subject of the subjective interpretation of the second law is the spatialized subject. This spatialized attitude underlies the conservative, and naïve character of modern physicists. In this framework, the concept of time is stripped of heterogeneity—the coexistence of qualitatively distinct elements—and physics proceeds toward a pragmatically oriented formalization. While such physics may attain increasing formal rigor, it also, as Husserl cautioned \cite{Husserl}, risks becoming ever more alienated from human meaning. In this trajectory, the subject plays no constitutive role in physics from the outset.

%By contrast, the essential agnostic attitude regards the unknowability inherent in physics as ontologically fundamental and epistemologically productive. It opens the way for a conception of time that retains its original heterogeneity, as emphasized by Bergson—a temporality in which qualitatively distinct elements coexist. This direction gestures toward a physics in which the subjective involvement of the scientist is no longer obscured but instead acknowledged as constitutive. In such a vision, the role of the subject in physics is no longer incidental but is actively and correctly integrated. The relation between physics and the human is thus not merely formal, but becomes a necessary and inseparable nexus of meaning, from which physics can continue to develop.

Of course, Bergson's philosophy and Prigogine's work do not completely overlap. Prigogine engaged with the concept of time within the realm of physics, whereas Bergson dealt with a broader critique of science, metaphysics, and philosophy, focusing on the fundamental errors that human intellect inevitably makes when attempting to conceptualize time. In this sense, Bergson and Prigogine operate discuss at different levels. Nevertheless, their projects share a common goal. Just as Bergson sought to restore the true nature of time, challenging both science and philosophy for reducing it to a static, discontinuous entity, Prigogine sought to restore time within physics itself. In the physics of being, dissipation and time were regarded as excessive, incomplete, and fundamentally negative concepts. Prigogine's work aimed to reposition them in their rightful place, restoring their intrinsic significance within physics.

\section{Summary}

In reflecting on Prigogine's scientific achievements alongside his philosophical insights, one gains a perspective not only on physics itself, but also on the underlying thought, tendencies, and strategies of the physicists who shape it. Physics, as it has developed, tends to formulate time in certain manner by extending particular attributes, and this can be categorized into four distinct conceptions of time. The systematization of these conceptions, as Prigogine suggests, reveals that they are deeply rooted in the worldview held by physicists. We found that they are, in fact, fundamentally divided into two branches, dynamics and thermodynamics, depending on how the notion of "dissipation" is positioned. It was shown that physicists tends to have consistently gravitated toward the dynamicist standpoint.

To clarify this, I have formulated into two attitude based on the agnosticism: the {\it naïve agnostic attitude} and the {\it essential agnostic attitude}. Building on this framework, I have sought to expand and elaborate on Prigogine's idea of the {\it temporalization of physics} by referring to the philosophies of Bergson, Deleuze, and Kimura, as well as the insights provided by the pitch drop experiment and the framework of intermediate asymptotics. This allows us to address two central questions: In what sense has physics been spatialized? And what could it mean for physics to be temporalized? We found that the physics has been spatialized so as that the heterogeneity between subject and object is eliminated. This tendency and strategy is the character of modern physics while Prigogine might have had an idea of another way, in which the heterogeneity is considered to be a positive attribute.

From this perspective, Prigogine's work, at least partially, can be seen as a form of cultural revolution mediated through natural science. His proposed new relationship between humans and nature resonates with various philosophical traditions \cite{Ende,Ushiki,Maruoka2017}. His analysis was highly influenced from Bergson and Deleuze
His analysis of time parallels Bergson's philosophy, expanding it into the realm of physics on a different conceptual level. This idea can potentially be further extended by Deleuze \cite{Deleuze1956,Deleuze1966}. The \changed{complicit relation} between humans and nature resonates with Uexküll's concept of Umwelt (the surrounding world) \cite{Uexkull} and Viktor von Weizsäcker's Gestaltkreis theory \cite{Weizsacker}. Similar insights can be found in Japanese psychiatrist Bin Kimura's work \cite{Kimura, Phillips,Fukao} and in Friston's Free-Energy Principle in cognitive science \cite{Friston}. The distinction between the two agnostic attitudes observed between physicists and Prigogine can potentially be related with the distinction between {\it mere coarse-graining} and {\it substantial coarse-graining}  \cite{Morita}. The complicit relation can be comparable to the {\it relational interpretation} by Rovelli \cite{Rovelli2019,Rovelli2022}. Exploring these directions could be future projects.

After all, Prigogine's contributions exhibit a fundamentally diverse nature, but this heterogeneity serves as direct evidence that his ideas remain, in some sense, still alive. While his vision has not been explicitly carried forward in contemporary physics, it is evident that his influence continues to unfold in various ways. As Prigogine and Bergson, some others emphasized and suggested, physics, nature, and humane are not {\it closed} systems---they remain {\it open}.

%% body of a paper

%% references without numbers
%\begin{references}
%\item
%\end{references}

%% references with numbers

\section{Acknowledegemnt}

The author wishes to thank  A. Koide, Y. Hirai, J. Aames,  K. Koseki, A. Takahashi, H. Saito, M. Murase, A. Belyi and the participants of the seminar of Prigogine for their fruitful discussion. The author wishes to thank K. Fukao, Y. Miwaki, Y. Maruhashi, W. Niwa and S. Shiba for the opportunity to have the discussion in the seminar of Aporia. The author particularly thanks Y. Maruhashi for consultations on Weizs\"{a}cker and Kimura. This work was motivated by the online seminar by the author for the Time and Contingency Research Group, which was organized by R. Ohmaya and I. Motoaki\cite{Maruoka2022}, by discussion with H. Ushiki and members in the seminar of Deleuze at Kyoto University, and by an encounter with B. Kimura. ChatGPT was used for writing assistance.

\end{document}